\documentclass[prl,10pt,superscriptaddress,twocolumn]{revtex4}

\usepackage[latin1]{inputenc}
\usepackage[english]{babel}
\usepackage[draft]{hyperref}
\usepackage{graphicx}
\usepackage{amsmath}
\usepackage{color}
\usepackage{float}
\usepackage{amssymb}
\usepackage{times}
\usepackage{amsmath}
\usepackage{dcolumn}
\usepackage{colordvi}
\usepackage{epsfig}

\begin{document}

\title{Quantum Metrology for Gravitational Wave Astronomy}

\author{Roman Schnabel*}
\affiliation{Albert-Einstein-Institut (AEI), Max-Planck-Institut f\"ur Gravitationsphysik and Leibniz Universit\"at Hannover, Callinstr. 38, 30167 Hannover, Germany}

\author{Nergis Mavalvala}
\affiliation{LIGO Laboratory, Massachusetts Institute of Technology,
Cambridge, MA, USA}

\author{David E. McClelland}
\affiliation{Department of Quantum Science, Research School of Physics and Engineering, The Australian National University, Canberra, 0200, Australia}

\author{Ping Koy Lam}
\affiliation{Department of Quantum Science, Research School of Physics and Engineering, The Australian National University, Canberra, 0200, Australia}

\date{Oct. 01, 2010}

\begin{abstract}
Einstein's General Theory of Relativity predicts that 
accelerating mass distributions produce gravitational radiation,
analogous to electromagnetic radiation from accelerating charges.
These gravitational waves have not been directly detected to date,
but are expected to open a new window to the Universe in the near future. 
Suitable telescopes are kilometre-scale laser interferometers measuring the distance between quasi free-falling mirrors. 
Recent advances in quantum metrology may now provide the required sensitivity boost. So-called squeezed light 
is able to quantum entangle the high-power laser
fields in the interferometer arms, and could play a key role in
the realization of gravitational wave astronomy.
\end{abstract}

\maketitle

When Galileo Galilei pointed his telescope towards the sky 400 years ago, he discovered events that had never been seen before. In subsequent centuries a variety of telescopes were invented, covering a large part of the electromagnetic spectrum. These telescopes enabled observations that now form the basis of our understanding of the origin and the evolution of the Universe. Einstein's General Theory of Relativity, quite often simply `General Relativity' (GR) \cite{Ein16} predicts the existence of a completely different kind of radiation, the so-called gravitational waves (GWs). Analogous to electromagnetic radiation from accelerations of charges, GWs are produced by accelerating mass distributions such as by supernova explosions or binary neutron stars that spiral into each other. GWs may also be emitted by objects that are electromagnetically dark, black holes, for example. Instruments that can directly observe GWs may well be able to ``light up'' the dark side of our Universe. The analysis of the waves' spectrum and their time-evolution will provide information about the nature of astrophysical and cosmological events that produced the waves.
So far, GWs have not been directly observed.

% Main conclusions 
Suitable telescopes for GW astronomy are kilometre-scale laser interferometers that measure the distance between quasi free-falling mirrors.  This measurement can be used to infer changes of space-time curvature. Current GW detectors are already able to measure extremely small changes of distance with strain sensitivity down to the order of 10$^{-22}$. However, quantum physics imposes a fundamental limit on measurement sensitivity, in particular in terms of photon counting noise. 
%, and jeopardizes the realization of GW astronomy. 
In the past, the GW signal with respect to the photon counting noise could only be increased by increasing the light power. Unfortunately, an increasing light power may introduce quantum radiation pressure noise at some stage but, in particular, increases the thermal load inside the detector and confuses the issue of an overall low noise concept. Squeezed light solves the problem of increasing the measurement sensitivity \textit{without} increasing the light power. The application of squeezed light represents in fact a quantum technology. Injected into an interferometer, it entangles the high-power laser fields in the interferometer arms. The photons detected at the interferometer output port are then no longer independent from each other any more resulting in a reduced, i.e.\ \textit{squeezed}, photon counting noise. Since the squeezed light technology does not build on an increase of light power, it keeps the thermal load constant and can conveniently be used in conjunction with other future technologies.  In particular, it can be combined with the cryogenic cooling of interferometer mirrors for reducing mirror surface Brownian motion. Future GW observatories might actually require squeezed laser light in order to make GW astronomy a reality. Recent progress in the generation of squeezed laser light has brought us to the point where quantum metrology will actually find its
first application.

\begin{figure}[ht]
  \vspace{-0mm}
  \includegraphics[width=8.6cm]{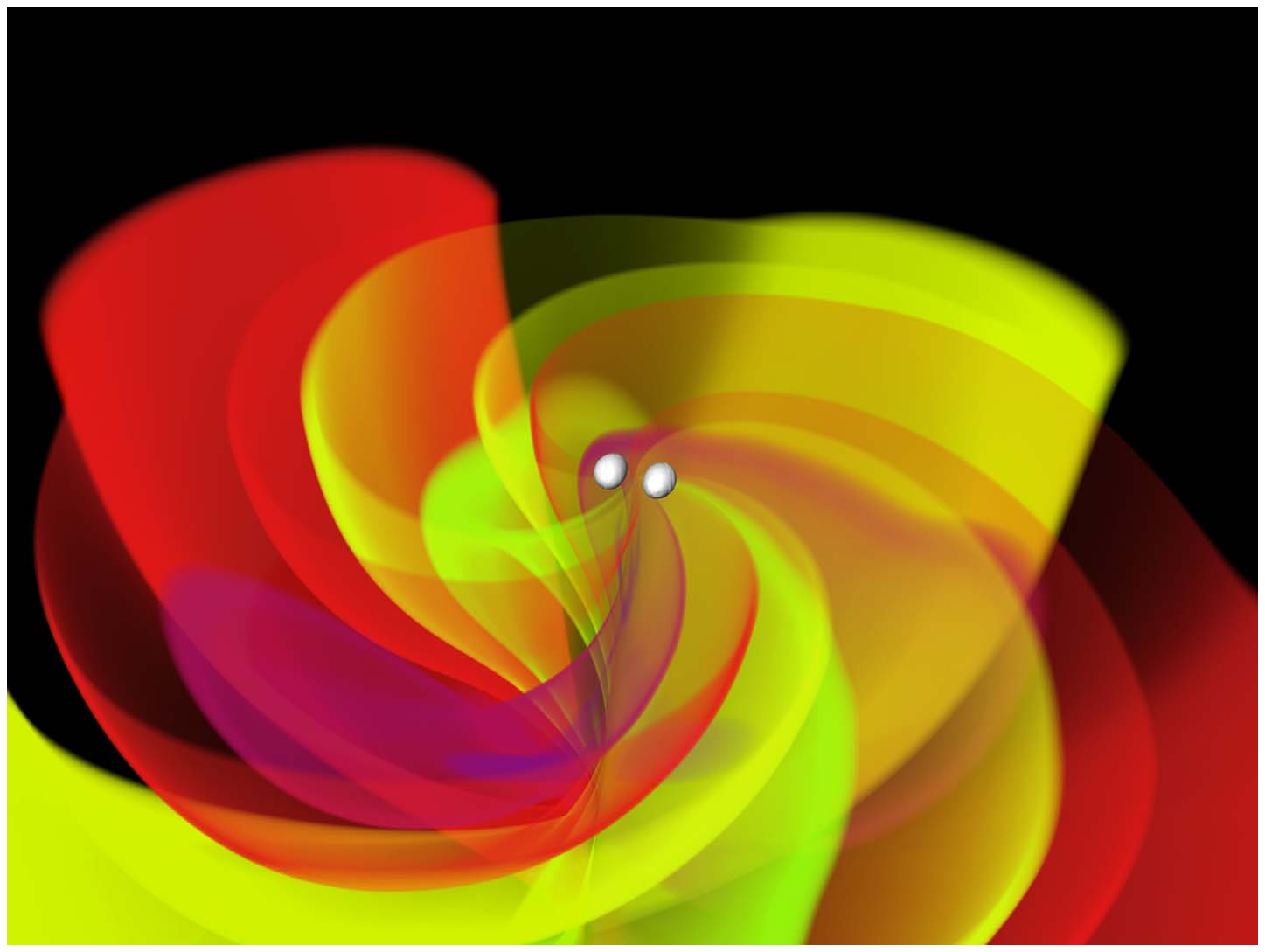} \vspace{-2mm}
\caption{\textbf{Merging neutron stars} \; Numerical relativity simulation of gravitational waves emitted from two neutron stars \cite{BGR08} which are about to merge in 4\,ms, taken from a movie \cite{aei-movie}. Shown is just the lower half of the sphere. The GW amplitude $h$ is colour-coded. At large distances to the stars, the wavelength is given by the distance of two wave fronts having the same colour. The time-resolved detection of these waves including the final merger phase could tell us what's inside neutron stars, i.e.\ their composition and the equation of state of matter at nuclear densities. By courtesy of L. Baiotti (AEI), R. Kaehler (AEI/ZIB), L. Rezzolla (AEI).
}
\label{fig1}
\end{figure}

\section{Gravitational waves} 

Gravitational waves are ripples in space-time, i.e.\ dynamic changes
in space curvature that propagate at the speed of light. According to GR they are transverse and quadrupolar in
nature, have two polarization states, and are extremely weak.  
GWs of detectable amplitude cannot be generated on Earth, but a variety of known
astrophysical and cosmological sources are predicted to emit
gravitational radiation that should reach the Earth with a
strength within reach \cite{SSc09,GWICweb}. 

Whilst GWs have not yet been directly observed their existence is beyond doubt.  A binary system of compact objects, such as neutron stars (as depicted in Figure 1) or black
holes, emit GWs at twice their orbital frequency . The energy carried
away by the GWs leads to a precisely predictable decay in the
orbital period of the binary. Hulse and Taylor verified this
mechanism for orbital decay to exquisite precision with observations
of the binary pulsar system PSR1913+16, \cite{WTa05}. Their
discovery is regarded as unequivocal, albeit indirect, proof of the existence of GWs that led to the 1993 Nobel Prize in Physics.

GWs from complex astrophysical sources carry a plethora of
information that will have a major impact on gravitational physics,
astrophysics and cosmology. GW signals are typically distinguished
in one of four broad and often overlapping classes
\cite{SSc09,GWICweb}, based on expected waveforms, and hence optimal
search techniques. They are: binary inspirals and mergers, burst
sources, periodic sources, and stochastic sources.  In the following we briefly review the physics and astrophysics that can be extracted from the observation of GWs emitted by these sources.

\textbf{Binary inspirals and mergers}\; 
The final stages of life of neutron star binaries will provide the
richest signals, see Fig.~\ref{fig1}. As the binary loses energy, the
orbital period decreases and enters the human audio frequency band.
After another $\approx 100$ cycles the stars merge in a catastrophic
explosion providing a GW burst signal of a few hundred Hertz up to a kiloHertz. 
The merger is expected to produce a black hole surrounded by a torus which will release a giant burst of gamma rays. Simultaneous
observation of GWs and gamma rays would confirm that the merger of
neutron stars is the engine of many of the observed short, hard
gamma ray bursts \cite{AbbottETAL08a}. Recent advances in numerical
relativity now make it possible to make predictions of the waveforms
generated around the merger \cite{BGR08}. Comparison with observed
waveforms will provide accurate tests of GR in the hitherto untested
strong-field regime. The imprint of tidal distortions on the GW
waveform from a binary system with at least one neutron star will
constrain the equation of state of the nuclear matter making up the
star. Independent of the nature of the binary, the final state of the merger will be a perturbed black hole, whose oscillation modes will decay in time producing more gravitational radiation. Such observations offer a striking confirmation of the existence of black holes. 

The famous ``no-hair'' theorem says that black holes are completely
characterized by their mass and angular momentum \cite{Chandrasekhar98}. Measuring the GWs emitted by black hole binary systems where the mass ratio of the components is large, the ``no-hair'' theorem
can be tested.  Direct observation of the gravitational waveforms from inspiralling black holes and neutron
stars can also provide the luminosity distance to the source without any complex calibrations
\cite{SSc09}. If, in addition, the redshift can be measured (via the identification of electro-magnetic counterparts), the
Hubble parameter \cite{Sch86}, the dark energy and dark matter content of the
Universe and the dark energy equation of state can be determined.

 \begin{figure}[ht]
  \vspace{-1mm}
  \includegraphics[width=8.6cm]{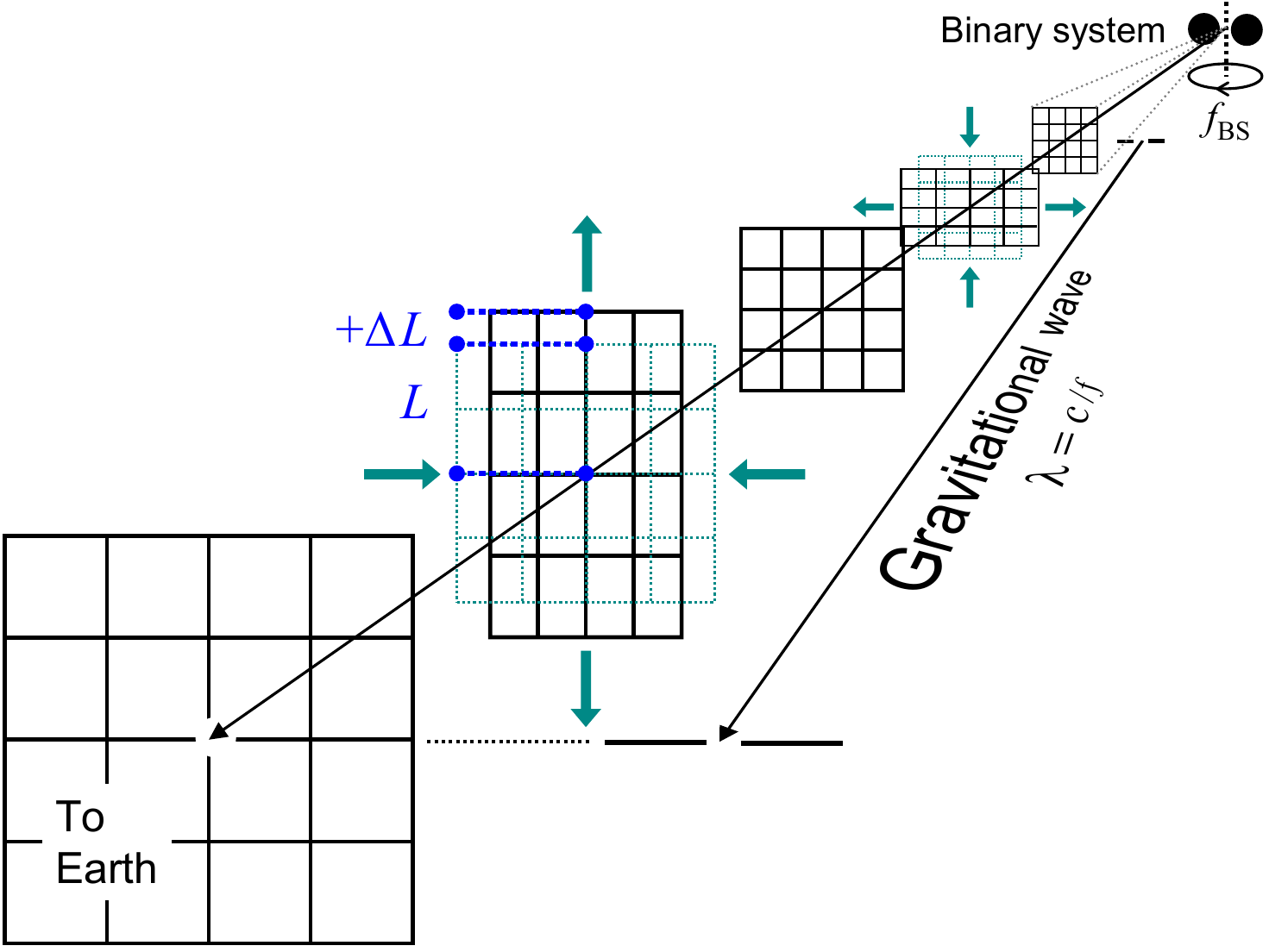}
  \vspace{-1mm}
\caption{
\textbf{Gravitational waves}\; GWs are dynamical deformations of space-time perpendicular to the direction of wave propagation. As a result, distances between free-falling test masses in a transverse plane will change with a strain $h = \Delta L / L$ with quantities defined as shown. For a black hole or neutron star binary system with orbital frequency $f_{\rm{BS}}$ distances will oscillate at twice that frequency $ f = f_{\rm{BS}}$. The wavelength of this oscillation is given by $\lambda =  c / f$ with $c$ the speed of light.
}
\label{fig2}
\end{figure}

\textbf{Burst sources}\; Burst sources refer to short-lived GW
transients, the main known candidates being core-collapse supernovae and collapses to black holes \cite{DFM01,BRe06}.
Observation of GWs will open a way to extract information
about the dynamics occurring in the core of the supernova, and
should complement and enhance the understanding gained from
electromagnetic observations.

\textbf{Periodic sources}\; Spinning compact objects will generate
periodic GW signals depending on the degree of non-axisymmetric deformations \cite{OGu69}
(departure from rotational symmetry is a necessary ingredient for generation
of quadrupolar moments). Detection of GWs from such sources will
confirm models of the underlying physics which might allow the
growth of a ``mountain'' on a neutron star. The lack of observation of
GWs from the Crab Pulsar at the sensitivity of current ground based
detectors has already constrained its deviation from rotational symmetry
\cite{AbbottETAL08a}. 
The distribution of neutron stars in the
Galaxy could be mapped out using GW observations. Spinning neutron
stars currently invisible on Earth could be detected via their GW
emission \cite{Abb09}.

\textbf{Stochastic sources}\; Stochastic sources have both
astrophysical and cosmological origins
\cite{Peebles93,Mag00}. The ``holy grail'' is the Big Bang itself. In principle, we
should be able to observe a relic background of  GWs from the very
early Universe, some time between 10$^{-18}$ seconds and 10$^{-9}$
seconds after the Big Bang, when light did not even exist. The
electromagnetic analogue of this radiation is the cosmic microwave
background, which gives information about conditions in the Universe
$385,000$ years after the Big Bang \cite{BenETAL03,SpeETAL03}.
Gravitational radiation is the only way to observe the conditions in
a much earlier epoch. Absence of a detectable stochastic background
signal in current GW detectors has constrained certain
models of the early Universe based on cosmic superstring population \cite{LIGO-Nat09}.

Of course the most tantalizing sources are those we do not yet know
exist. The opening of every major new electro-magnetic window to the
Universe has revealed major surprises that have revolutionized our understanding of the Universe. 
Observing the Universe with an entirely new messenger will very likely continue this tradition.

\textbf{Frequencies of GWs}\;  GW astronomy targets phenomena that involve astronomically large masses in acceleration.  This, in turn, leads to the
expectation that GW emission frequencies will be low, typically
below a few tens of kiloHertz. A black hole binary system, for
example, has to have an orbital period of just $0.02$\,s in order to
produce GWs at $f$=100\,Hz (Fig.~\ref{fig2}).
Supernova explosions are expected to have a broad spectral emission, with components that may reach kiloHertz
frequencies. However, the strongest detectable GWs are expected at
lower frequencies, all the way down to the millihertz or even nanohertz regime.

\textbf{Strength of GWs}\; Gravitational waves that reach the Earth are extremely weak. For example, the merger of two neutron stars at the other end of our galaxy ($D \approx $ 50,000 light years away) would produce a GW strain amplitude of about $h \approx 10^{-19}$ \cite{SSc09}. The same source at the distance of about 60 million light years, where the Virgo cluster which comprises up to 2000 Galaxies are located, would result in a corresponding strain amplitude of only $h \approx 10^{-22}$. 
With the sophisticated technology now
available, {such tiny strains of space-time can be detected and} it is very probable that there will be numerous direct
detections in the coming decade.

\section{Gravitational wave detection}  

\begin{figure}[ht]
  \vspace{-1mm}
  \includegraphics[width=8.6cm]{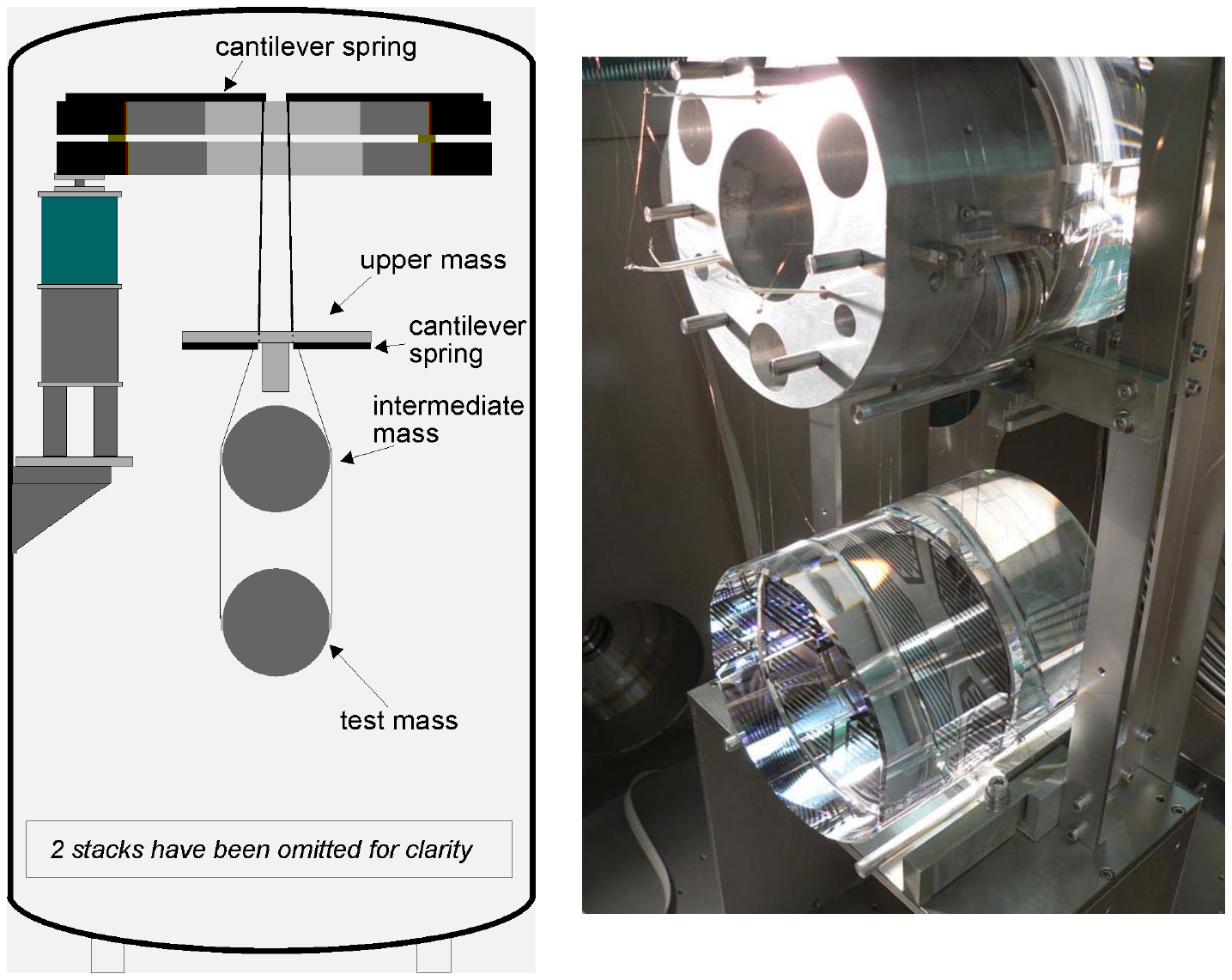}
  \vspace{-1mm}
\caption{{\textbf{Michelson interferometer} Continuous wave laser light is split into two beams traveling in orthogonal directions. Both beams are reflected back towards the central beam splitter. GWs change the optical path length difference, and thus the interference at the beam splitter and the light power directed towards the photo diode.
A GW at frequency $f$ reveals itself as a light power modulation at the same frequency. }
}
\label{fig3}
\end{figure}

GWs stretch and compress the spacetime transverse to their direction of
propagation. If the wave were incident on a ring of {free} test masses {in space}, in
each half cycle of the wave, the ring would distort into an ellipse,
as shown in Fig.~\ref{fig3}. If the test masses were mirrors, one could reflect laser light off them, and observe this GW
induced stretching and compressing of spacetime by measuring the light
travel time. This is, in fact, the principle that interferometric GW detectors are based on.  An overview of  the history of detectors based on an alternative measurement scheme can be found in BOX\,1. 

\textit{
~\\
\textbf{BOX 1} \\
\underline{Past and present GW detectors}\\
The first experimental attempt to directly measure GWs started in the 1960s \cite{Web60}. The detection principle was based on the GW-induced resonant excitation of vibrational modes of metal cylinders. Cryogenically cooled devices reached strain sensitivities of about $h = 10^{-18}$  around a kilohertz, over a band width of a few Hz, in the 1990's and have been further improved since then \cite{MPT87,CerdonioETAL97,JBZ00,AstoneETAL03}. 
Today, the most sensitive instruments are laser interferometers with kilometre size arm lengths.
In the past decade a global network of GW detectors has been realised.  The Japanese \textit{TAMA} project built a 300\,m interferometer outside Tokyo, Japan \cite{Tak04}; the British-German \textit{GEO} project built a 600\,m interferometer near Hannover, Germany \cite{Lue06,Willke06}, see Fig.~\ref{fig5}; the US-American \textit{LIGO} project built two 4\,km and a 2\,km interferometer on sites in Washington and Louisiana \cite{Abramovici92,AbbottETAL09ligo} and the European Gravitational Observatory maintains the 3 km-long interferometer \textit{Virgo} near Pisa, Italy \cite{Acernese08}. These detectors target the GW frequency band from 10\,Hz to 10\,kHz. Currently, the most sensitive detector, LIGO, has achieved a root-mean-square (rms) strain noise of $3\times10^{-22}$ in its most sensitive band from 100\,Hz to 200\,Hz \cite{AbbottETAL09ligo} thereby reaching its design sensitivity at these frequencies. At such a high sensitivity, detection of GWs is in principle possible.  However, an improvement in the sensitivity of gravitational wave detectors by about a factor of a hundred is required for gravitational wave astronomy.\\
~\\
}

The enormous difficulty of GW detection arises because GWs are expected to be
extremely weak when they finally reached the earth. The amount by which a distance $L$ would shrink
or stretch due to a GW is proportional to the wave's amplitude $h$, i.e.\
$\Delta L = h\,L$. Recalling that we expect strain amplitudes of
$10^{-22}$, we are faced with the prospect of measuring changes in
separation of $10^{-18}$~m even for a one-kilometre interferometer.

The intrepid GW detector designer thus faces two categorical
challenges. First, how to keep the test masses so still that they respond
only to a passing gravitational wave, rather than to local perturbations? This
isolation problem is addressed by techniques of vibration isolation
and material engineering and has to be optimized for the targeted
frequency spectrum. Second, how to measure relative displacements with
sufficient precision? This measurement problem is tackled by adopting
advanced techniques in optical interferometry, control theory, and quantum metrology. Let us tackle
the question of the mechanical design for an earth-based test mass of
spacetime first, followed by a discussion of metrology which
launches us into the optical design of the instrument.

\begin{figure}[ht]
  \vspace{-1mm}
  \includegraphics[width=8.6cm]{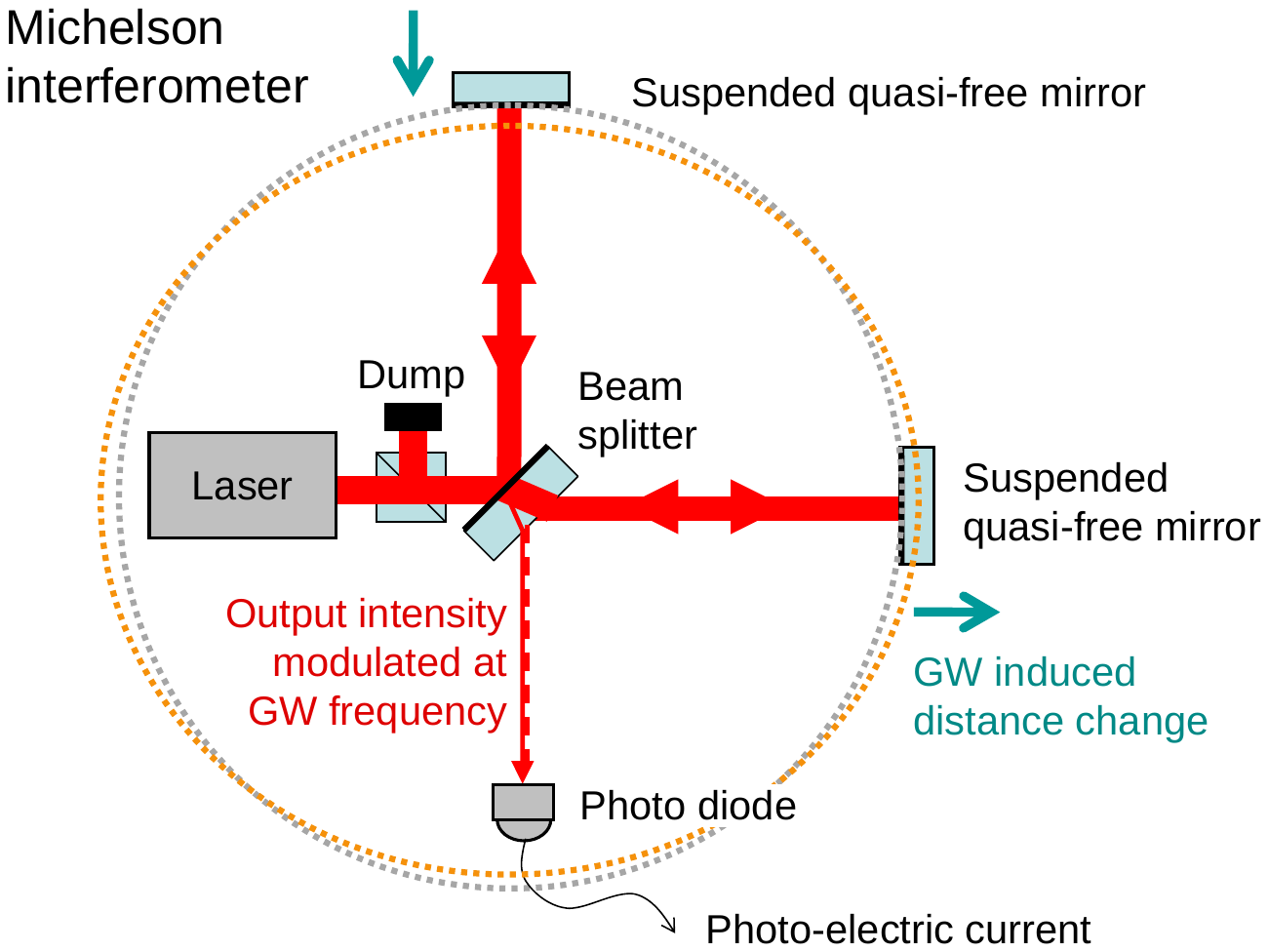}
  \vspace{-1mm}
\caption{\textbf{Quasi-free falling test mass} \; A GW detector requires laser mirrors as test masses in space-time. Left: Due to the in-vacuum threefold pendulum suspension, the bottom mirror is quasi-free falling in direction of laser beam propagation and highly decoupled from the environment. Right: Mirrors of today's GW detectors are made of dielectrically coated low absorption fused silica. Actuated electro-static forces between the mirror and a reaction mass placed 3\,mm behind allow for a stabilization of the interferometer close to its dark fringe. By courtesy of the AEI and the GEO\,600 collaboration.} \label{fig4}
\end{figure}

\begin{figure}[ht]
  \vspace{-1mm}
  \includegraphics[width=8.6cm]{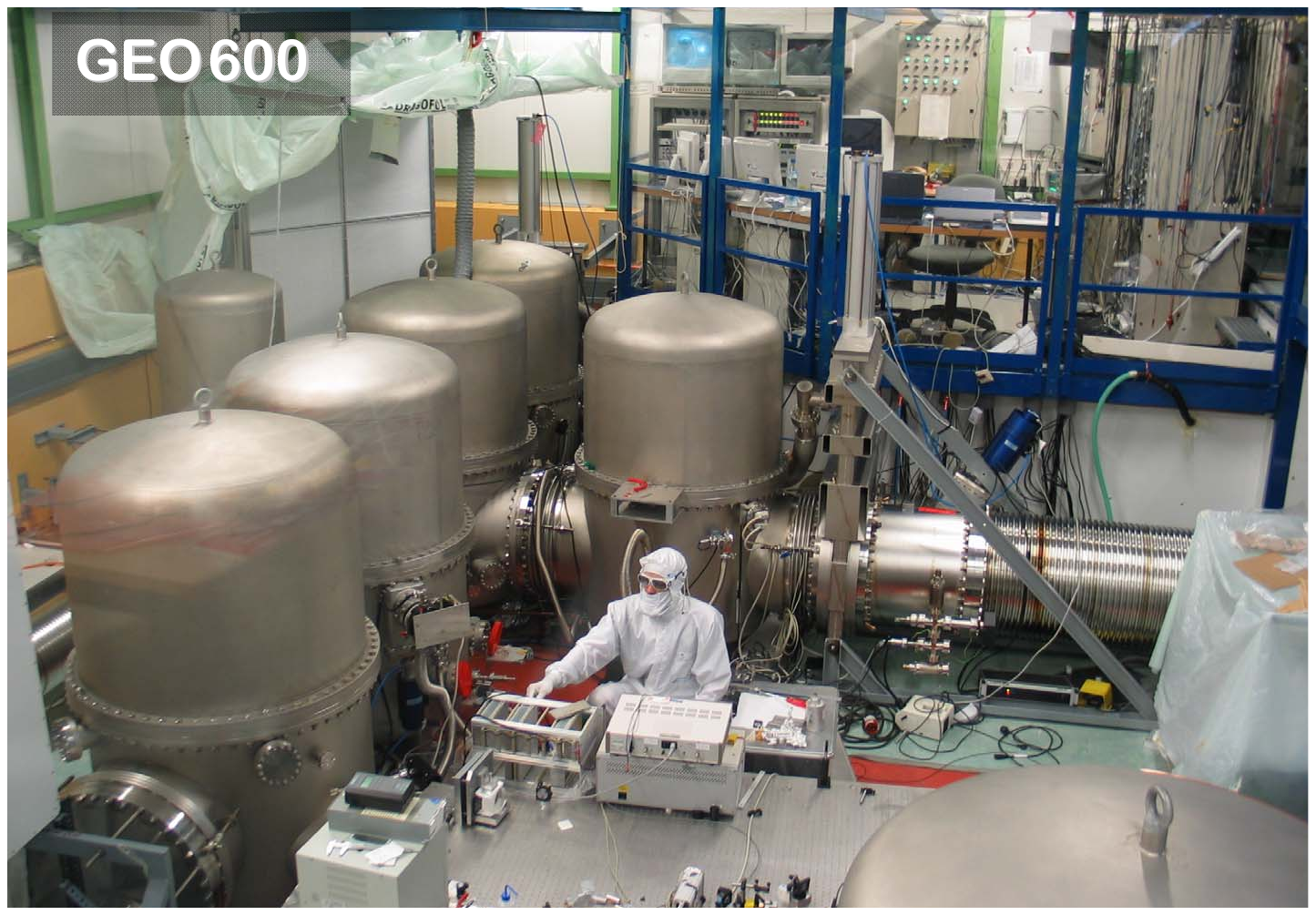}
  \vspace{-1mm}
\caption{\textbf{GEO\,600}\; View into the central building of the British-German GW detector located close to Hannover, Germany. The vacuum chambers contain the suspended beam splitter, power- and signal recycling mirrors, additional input and output optics as well as mirrors to realize a double pass of the laser light through the 600\,m long interferometer arms. By courtesy of the AEI.}
\label{fig5}
\end{figure}

The mirrors of interferometric GW
detectors are designed to be quasi-free falling in the directions of propagation of the
laser beams, thereby acting as test masses that probe spacetime.
This is achieved by suspending the mirrors as sophisticated
pendulums in vacuum chambers, as shown in Fig.~\ref{fig4}. Above the
pendulum's resonant frequency, typically around 1\,Hz, the suspension
isolates the mirror from vibrations of the ground and the structures
on which it is mounted making it ``quasi-free''. The targeted detection band of earth-based detectors is therefore restricted to the audio-band (to frequencies above $\approx$10\,Hz). At lower frequencies disturbances from the environment are too high, at higher frequencies no strong GW signals are expected, see previous section. 

The mirrors and their suspensions are built from materials having
exquisitely high mechanical quality factors. This helps to
concentrate the thermal energy that causes displacements of the
mirror surface into well-defined vibrational frequency modes. At these
particular frequencies, no GWs can be detected. The vibrational modes are therefore designed
to be outside of the detection band for the most part. Ultimately
cryogenic cooling of mirror suspensions may have to be used to
further reduce thermally excited mirror displacement noise, such as those originated from
Brownian motion. The first cryogenic interferometric GW detector
prototype facilities have been recently realized \cite{AraiETAL09}.

A Michelson interferometer -- similar to the
one used in the Michelson-Morley experiment, which famously
established that the speed of light was a directionally invariant
\cite{MichelsonMorley1887} -- is ideally
suited to measure the relative light travel time in two
orthogonal directions (Fig.~\ref{fig3}). In a Michelson interferometer, laser light
is incident on a beam splitter that reflects half the light, and
transmits the other half. Each light beam travels some distance
before it is reflected by a mirror back towards the beam splitter where the two beams interfere.
The interference provides an output beam whose power carries information
about the path difference, and gravitational wave signals are detected as variations in the light power.

It is at this point that quantum physics enters the concept of gravitational wave detection. First of all, the light's energy can only be absorbed in discrete quanta (photons), resulting in photon counting noise, or shot-noise. The GW signal to shot-noise ratio can in fact be improved by detecting more photons. Shot-noise is proportional to the square root of the number of photons detected, while the mirror displacement signal is directly proportional to the laser
power. Consequently, GW detectors use high-power laser systems and optical resonators to
maximize their shot-noise-limited sensitivity (For further details please refer to BOX\,2).

\textit{
~\\
\textbf{BOX 2} \\
\underline{Signal to shot-noise improvement by classical means}\\
In the past decades several advanced interferometer techniques based on optical resonators were invented to further increase the signal-to-shot-noise ratio in GW detectors. Generally, GW detectors are operated close to a dark fringe, i.e.\ the steady state mirror separation is arranged for nearly perfect destructive interference on the photo diode. This operation point not only cancels common mode noise such as laser noise but also maximizes the signal to shot-noise ratio. Furthermore, since most of the laser power is reflected back towards the laser, a partially reflecting mirror placed between the laser system and the beam splitter resonantly enhances the light power inside the interferometer.  This  technique is known as power-recycling \cite{DHKHFMW83pr}. Similarly, a partially reflecting mirror placed between the output port of the beam splitter and the photodiode can be used to resonantly enhance the GW signal; this is known as signal recycling \cite{Mee88sr}.
Finally, two partially reflecting mirrors placed near the beam splitter turn the Michelson interferometer arms into kilometre scale Fabry-Perot cavities to increase the phase sensitivity of the interferometer by causing the light to interfere multiple times with itself. 
All these techniques are classical techniques that maximize the signal-to-shot-noise ratio. At frequencies above a few hundred Hertz, shot-noise is still the limiting noise source for
gravitational wave detectors.\\
~\\
}

Fundamentally, there is a second way how the quantum noise of light disturbs a GW detector. The shot-noise inside the interferometer produces a fluctuating radiation pressure force on the test mass mirrors. The mirrors are randomly displaced by the light, an effect that cannot be distinguished from a gravitational wave signal.  This is called quantum radiation pressure noise \cite{Cav80}. To reduce this effect, modern GW detectors use test masses of up to 10\,kg.  As a consequence, radiation pressure noise has not been experimentally observed to date.  This situation, however, may change with increasing laser power and is envisioned in the next generation of GW detectors.

The design of second generation GW detectors is more or less completed. These so-called \textit{Advanced} detectors will replace the existing interferometers aiming for a ten-times increased sensitivity \cite{Har10,Wei02,AdvLIGO06}. New laser systems will provide up to 200\,W of single mode optical power \cite{Frede05} to reduce quantum shot-noise yielding a light power of almost a megawatt in the interferometer arm resonators. Larger, 40\,kg test mass mirrors will replace the existing ones to keep radiation pressure noise low and to allow for larger beam radii to reduce the noise effect of mirror Brownian motion. Cryogenic cooling of test mass mirrors is another advanced technology that is planned to be implemented in a Japanese detector \cite{Kur06,AraiETAL09}. At very cold temperatures Brownian motion and other forms of thermally excited mirror surface motions (thermal noise) can be significantly reduced.

Theoretical modelling of GW sources and estimations of GW event rates \cite{SSc09} suggest that real GW astronomy with detections on a daily basis with high signal-to-noise ratios require another ten-fold sensitivity increase for ground-based observatories at frequencies down to a few Hertz.  At even lower frequencies noise on earth is too high and space-based observatories, such as LISA \cite{LISAweb}, are required, targeting a frequency spectrum from 10$^{-4}$\,Hz to about 1\,Hz. Above 1\,Hz, the \textit{Einstein Telescope} \cite{PunturoETAL10,ETweb} is an on-going European design study project for a third-generation ground-based gravitational wave detector. An important issue will be the further reduction of the shot-noise (quantum measurement noise), radiation pressure noise acting on the mirrors (quantum back-action noise) and thermal noise. The required reduction of these noise sources poses serious technical challenges. For example, increasing the light power in the interferometer arms will lead to additional absorption and heating of the mirrors.  Higher light power will also increase radiation pressure noise.  The only classical approach to mitigate noise is, therefore, to use even more massive mirrors.  An increased mirror thickness will again lead to increased absorption and heating, making cryogenic cooling of the mirrors impractical.  A quantum metrological approach is able to break this vicious circle. In the next section we will see that squeezed laser light is able to achieve a quantum noise reduction without increasing the light power in a GW  detector.\\ 
~\\

\section{Quantum metrology}  

``Metrology'' is the science of measurement. At first glance,  
quantum physics imposes a fundamental limit on metrology, and thus imposes a corresponding limit on the sensitivity of GW detectors.  A fundamental problem in optical interferometry is
the stochastic distribution of photons arriving at the photodiodes.  This statistical fluctuations obscure the tiny power variations caused by GW signals. Fortunately, quantum physics also provides a solution to this problem via the concept of quantum entanglement.

``Quantum metrology'' uses quantum entanglement to improve the
measurement precision beyond the limit set by measurement counting
noise. The first such proposal was made by C.M. Caves in 1981 when
he suggested the use of squeezed states of light as an (additional) input for
laser interferometric GW detectors \cite{Cav81}. 
Caves's initial proposal was motivated by
the limited laser power available at the time. Indeed, squeezed
states allow for improvement in the sensitivity of a quantum noise
limited interferometer without increasing the circulating laser
power.

\begin{figure}[ht]
  \vspace{-1mm}
  \includegraphics[width=8.6cm]{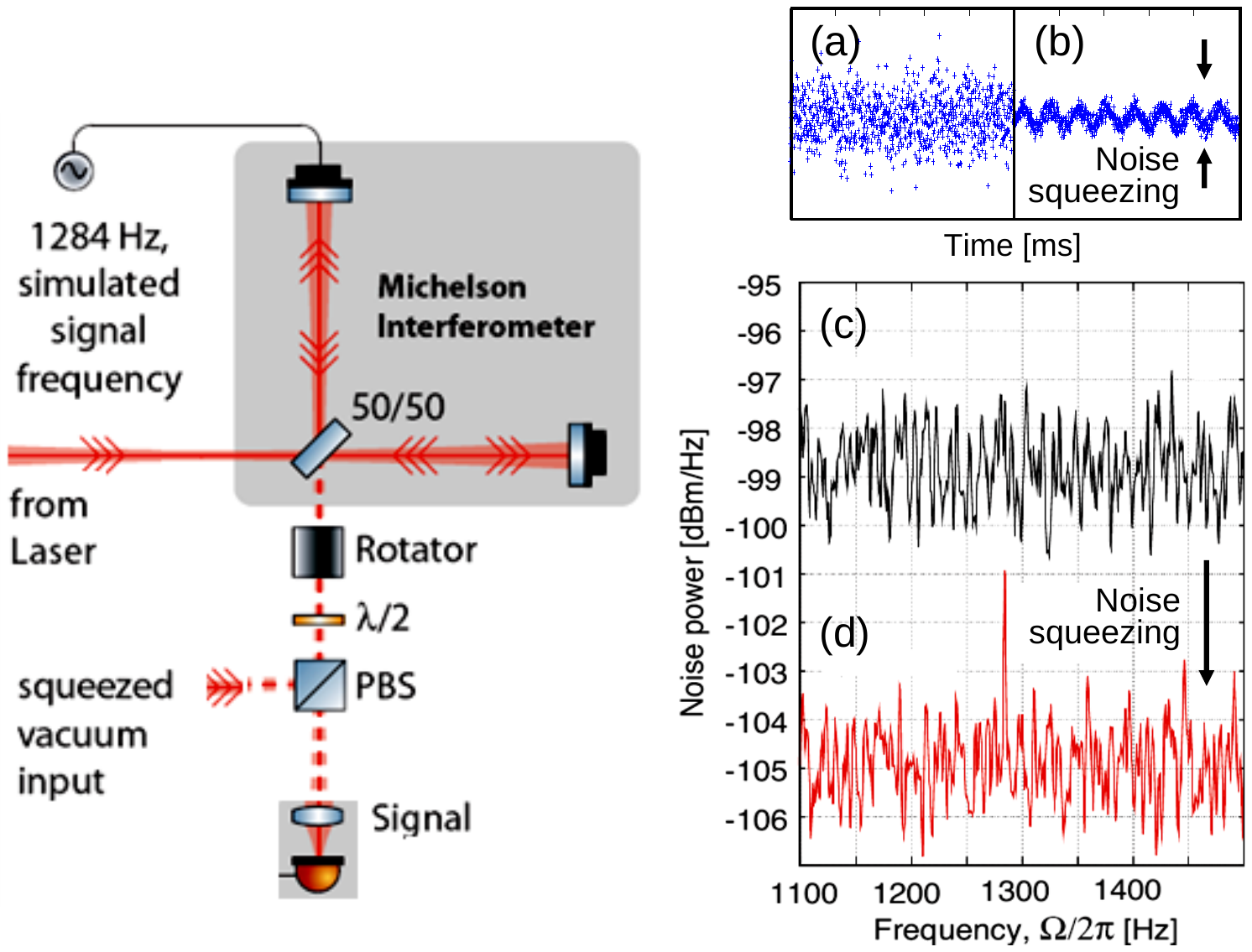}
  \vspace{-1mm}
\caption{{\textbf{Squeezed light enhanced metrology} \; (a) For large photon numbers $N$, squeezed light shows a photon counting statistic with a standard deviation smaller than $\pm \sqrt{N}$. In all panels, (i) correspond to shot-noise and (ii) to 6\,dB squeezed noise. (b) A squeezed vacuum beam is injected into the dark signal port of a Michelson interferometer, in addition to the conventional bright laser input. The squeezed beam leads to path entanglement of the light fields in the two arms and to an improved signal to noise ratio, as shown on the right. Without squeezing, the optical path length modulation at 1284\,Hz is neither visible in the time series of the photo-electron current (c, simulation by B. Hage, AEI) nor in its noise power spectrum (d, measurement, courtesy of H. Vahlbruch, AEI \cite{VahlbDiss}). In (c) as well as in (d), the signal is clearly visible when squeezing is applied (ii).} 
}
\label{fig6}
\end{figure}

Squeezed states \cite{Yue76a,Wal83,BSM97,Dod02} belong to the class of so-called \textit{nonclassical} states of light. 
Generally, nonclassical states are those that cannot be described by a classical (positive valued) probability distribution using the coherent states as a basis (the $P$-representation) \cite{GerryKnight04}. Let us first consider the coherent states. If light in a coherent state is absorbed by a photodiode, mutually independent photon `clicks' (in terms of photo-electrons) are recorded, a process that is described by a Poissonian counting statistics. Due to quantum mechanics, every individual `click' is not predictable, but rather the result of a truly random process. If the number of photons per time interval is large ($N\!>\!\!>\!1$), its standard deviation is given by $\sqrt{N}$, see Fig.~\ref{fig6}\,a\,(i). This uncertainty gives rise to \textit{shot-noise}. For a \textit{squeezed} light beam, the detection of photons are not time-independent but instead contains quantum correlations.  Nevertheless, the photon statistics still cannot be predicted by some external clock.   They instead show auto-correlations that give rise to a reduced standard deviation, as shown in Fig.~\ref{fig6}\,a\,(ii). The correlations might be described in the following way. Whenever the quantum statistics might drive the actual photon number above the average value $N$, a similar number of photons destructively interferes with the main body of photons providing a (partial) compensation of the fluctuation. These quantum correlations \textit{squeeze} the interferometer's shot-noise below its natural value. Another complementary way of describing the properties of squeezed states is based on the phase space quasi-probability distribution using the amplitude and phase quadratures of a light wave (the Wigner function) \cite{Wal83,GerryKnight04}.

A squeezed state that contains only quantum-correlated photons with no coherent amplitude is called a \textit{squeezed vacuum state} \cite{GerryKnight04}. If such a state is overlapped with a coherent laser beam on a semi-transparent beam splitter, two beam splitter outputs are generated which are quantum correlated. As a consequence, the overall (bi-partite) quantum state cannot be written in terms of products of the two beam splitter output states. Such a quantum state is called non-separable or \textit{entangled}. This is exactly what happens if a squeezed state is injected into the signal output port of a laser interferometer for GW detection (Fig.~\ref{fig6}\,b). The two \textit{high-power} light fields in the interferometer arms get entangled and the light's quantum fluctuations in the two arms are correlated with each other. Although the fluctuations are not predictable from the outside, they provide an improved signal-to-noise ratio in the interferometer. Recall that an interferometer measures the optical path length change in one interferometer arm with respect to the other arm. If the quantum noise in the two arms is correlated it will cancel out. 
This entanglement interpretation was not discussed in the initial proposal by Caves.  Nevertheless, it shows that the application of squeezed states in interferometers is a real application of quantum metrology by its very own definition. The entanglement produced by splitting a squeezed state at a semi-transparent beam splitter was tomographically characterized and quantified in \cite{DHFFS07}. Fig.~\ref{fig6}\,c shows a simulated signal from a photodiode, without (i) and with (ii) \textit{squeezing}. The tiny modulation in the interferometer's output light due to the (simulated) passing GW is visible only with the improved signal-to-noise ratio. Fig.~\ref{fig6}\,d shows the analogue in frequency space, i.e.\ after a Fourier transform of the photo current was applied.

The above paragraph shows that squeezed states can be conveniently combined with the extremely
high photon numbers of coherent light to improve a laser interferometer, as proposed in \cite{Cav81}
and shown in Fig.~\ref{fig6}\,b. In fact, the stronger the squeezing
factor \cite{Wal83,GerryKnight04} the greater the path entanglement and the signal-to-noise
improvement. Very strong path entanglement is present in interferometers using so-called NOON-states instead of squeezed states.  
NOON states are another class of nonclassical states \cite{GerryKnight04,HBu93,WPAUGZ04,AAS10}.   
Unfortunately, the strong
entanglement of a NOON state is extremely fragile, in particular if
$N$ is large. Very recently a NOON-state with $N=5$ photons was
demonstrated \cite{AAS10}. However, gravitational wave detectors use
coherent high-power laser light with $N \approx 10^{23}$ photons per
second. An improvement by use of NOON states is, therefore, far out
of reach.

Shortly after Caves proposed squeezed states of light for laser
interferometers in 1981, the first experimental demonstration of squeezed
light \cite{SHYMV85} and proof of principle demonstrations of
quantum metrology were achieved \cite{XWK87,GSYL87}. In parallel, it was theoretically discovered that squeezed states offer even more advances in metrology than `just' reducing the quantum shot-noise. 
From the early days of quantum physics, when fundamental
aspects of the measurement process were discussed, it was clear that,
in general, a measurement disturbs the system to be measured
\cite{Braginsky95}. The measurement of quantity $A$ (say
a position of a mirror) increases the uncertainty of the
non-commuting quantity $B$ (say the mirror's momentum). Both
observables are linked by a Heisenberg Uncertainty relation. For
repeated measurements of $A$, the increased uncertainty in $B$
disturbs the measurement of $A$ at later times. This is referred to
as \textit{quantum back-action noise}. Here, the back-action arises
from the fluctuating radiation pressure due to the reflected light \cite{Cav80}.
It is significant if the mirror's mass is low and a
large photon number is reflected. In the 1970s, ideas were developed
that showed how, in principle, back-action noise for continuous
measurements can be avoided. Such schemes were called
quantum-non-demolition (QND) measurements \cite{TDCZS78,BKh96}. 
However, for laser interferometric GW detectors
using \textit{quasi-free} falling mirrors it remained unclear if QND schemes
exist. In \cite{Cav80,Cav81} it was concluded that back-action noise
of a free mass position measurement can in principle not be avoided and, together with photon counting noise, defines a \textsl{standard
quantum limit} (SQL). In \cite{Yue83,Unr83} it was argued, however, that measurements below the
SQL of a free mass are indeed possible. The discussion
remained controversial \cite{Cav85} until Jaekel and Reynaud
\cite{JRe90} were able to convincingly show that the cleverly
arranged squeezed states in a GW detector can simultaneously reduce
the shot-noise and the radiation pressure noise, by almost arbitrary
amounts (as long as most of the photons belong to the light's coherent displacement). 
For a summary of QND techniques for free mass
position measurements we refer to \cite{KLMTV01}.

So far no experiment has achieved a position measurement with
sensitivity even at, let alone below, its standard quantum limit.
Eventually this will be achieved, possibly first in future
gravitational wave detectors.  Advanced detectors are in
fact designed to have a sensitivity at or just below their SQLs. Once the SQL is reached a new level of
quantum metrology is achieved, because the position-momentum
uncertainty of the mirror becomes correlated with the quadrature
uncertainty of the reflected optical field. In this way, entanglement
between the mechanical and the optical system can be observed
\cite{vitaliPRL2007}. This is all the more remarkable from the perspective of GW detectors
since we are talking about mirrors with masses of 40 kg, planned for
the upcoming improvement to LIGO - the Advanced LIGO \cite{Wei02}. Eventually,
even two such mirrors might be projected via entanglement swapping
\cite{pirandolaPRL2006} into an entangled state \cite{Helge08}.
Obviously quantum metrology opens the possibility for further
studies of the peculiarities of quantum physics at a macroscopic scale.

\section{Squeezed light for gravitational wave astronomy}   

Laser interferometers for GW astronomy are facing extreme sensitivity requirements that can only be achieved if all available
tools, inclusive of quantum metrology, are combined in an elaborate measurement device.  More recently, squeezed light was also suggested as a resource for quantum information processing \cite{YSh78,YHa86,STe87,BLo05}. Since then, squeezed light has been central to various proof-of-principle demonstrations, such as quantum teleportation \cite{FSBFKP98,BTBSRBSL03} and the production of optical ``Schr\"odinger cat'' states for quantum computing and fundamental research on quantum physics \cite{OTLG06,NNHMP06}.

Squeezed light must be generated in a nonlinear
interaction. Squeezed light was first produced in 1985 by
Slusher et al.\ using four-wave-mixing in Na atoms in an optical
cavity \cite{SHYMV85}. Shortly after, squeezed light was also
generated by four-wave-mixing in an optical fibre \cite{SLPDW86}
and by parametric down-conversion in an optical cavity containing a
second order non-linear material \cite{WKHW86}. In these early day
experiments, squeezing of a few percent to 2  to 3\,dB were routinely observed (For an
overview of earlier experiments and squeezed light generation in
the continuous-wave as well as pulsed regime please refer to Ref. \cite{BachorRalph2004}).

\begin{figure}[ht]
  \vspace{-0mm}
  \includegraphics[width=8.6cm]{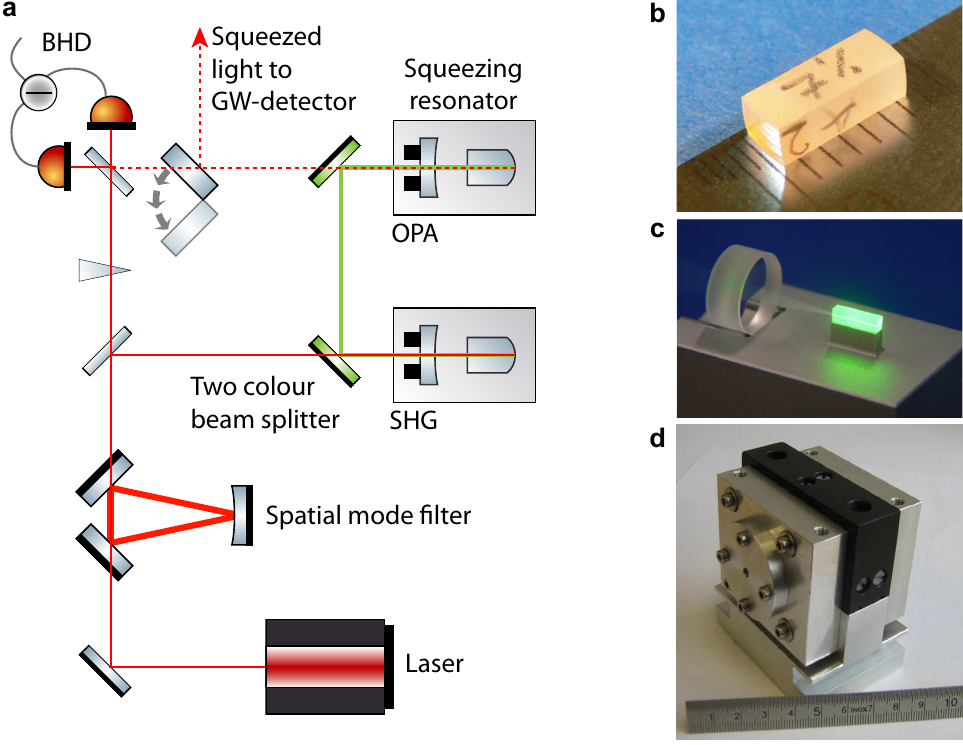}
  \vspace{-1mm}
\caption{{\textbf{Generation of squeezed light} (a) A continuous-wave laser beam at the GW detector wavelength is first spatially filtered and then up-converted to a field at half the wavelength (second harmonic generation, SHG). That beam is then mode-matched into the `squeezing resonator' in which a tiny fraction of the up-converted photons are spontaneously down-converted by optical parametric amplification (OPA) producing a squeezed vacuum state. The squeezing factor is validated by a balanced homodyne detector (BHD). SHG as well as OPA are realized by a nonlinear crystal (b), here a 6\,mm long MgO:LiNbO$_3$ crystal, inside an optical resonator (c) formed by an external cavity mirror and the dielectrically coated crystal back surface. The two nonlinear resonators may be constructed in an identical way and are put into temperature stabilized housings (d).} 
}
\label{fig7}
\end{figure}

GW detectors are operated with high-power, quasi-monochromatic
continuous-wave laser light with an almost Fourier-limited spatial
distribution of a Gaussian TEM$_{00}$ mode. For a nonclassical
sensitivity improvement, squeezed light in exactly the same
spatio-temporal mode must be generated and mode-matched into the
output port of the interferometer \cite{Cav81}, providing
interference with the high-power coherent laser beam at the
interferometer's central beam splitter. High-power lasers for GW
astronomy are based on optically pumped solid-state crystals in
resonators \cite{Frede05}, suggestive of a similar configuration for
a ``squeezed light resonator''. Fig.~\ref{fig7}\,(a) shows a
schematic setup for generation of squeezed light that is built upon one of the very first squeezing experiments \cite{WKHW86}, a setup that has been used in many experiments thereafter
\cite{FSBFKP98,BTBSRBSL03,SLMS98,LRBMBG99}. The setup uses a solid
state laser similar to those used as master lasers in high-power
systems. After spatial mode filtering, second harmonic
generation (SHG) in an optical cavity containing a second-order
nonlinear crystal is applied to produce laser light at twice the
optical frequency. The second harmonic light is then mode-matched
into the squeezing resonator to pump a degenerate optical parametric amplifier.
 
Fig.~\ref{fig7}\,(b-d) show photographs of the nonlinear crystal, the optical arrangement and the housing of a squeezing resonator. The crystal is temperature
stabilized at its phase matching temperature. At this temperature the first-order dielectric
polarization of the birefringent crystal material with respect to the pump is
optimally overlapped with the second-order dielectric polarization of
the resonator mode at the fundamental laser frequency. This ensures
a high energy transfer from the pump field to the fundamental
Gaussian TEM$_{00}$ resonator mode, i.e.\, efficient parametric down
conversion.

Initially, the resonator mode is not excited by photons
around the fundamental frequency, i.e.\, it is in its ground state,
characterized by vacuum fluctuations due to the zero point energy
\cite{GerryKnight04}. Note that the process is typically operated
\textit{below} oscillation threshold in order to reduce phase noise coupling
from the pump \cite{RDr89}.  This setup produces a squeezed vacuum
state \cite{GerryKnight04}. The down-converted photon pairs leaving the squeezing
resonator exhibit quantum correlations which give rise to a squeezed
photon counting noise when overlapped with a bright coherent local
oscillator beam. The squeezed field is detected by interfering it
with a coherent local oscillator beam, either in a balanced homodyne
detector (BHD), see Fig.~\ref{fig7}\,(a), or when injected into a GW
detector and detected with a local oscillator from the GW detector
along with an interferometric phase signal, see Fig.~\ref{fig6}. The closer
the squeezing resonator is operated to its oscillation threshold,
and the lower the optical loss on down-converted photon pairs, the
greater the squeeze factor is. For instance, the observation of a
squeezing factor of 2 is only possible if the
overall optical loss is less than 50\% \cite{BachorRalph2004}. A
90\% nonclassical noise reduction, i.e.\ a squeezing factor of 10, or
10\,dB already limits the allowed optical loss to less
than 10\%.

Although squeezed light was demonstrated in the 1980s
shortly after the first applications were
proposed \cite{SHYMV85,SLPDW86,WKHW86}, several important challenges pertaining to the application of
squeezed states to GW detectors remained unsolved until recently.

First, squeezing has always been demonstrated at Megahertz
frequencies, where technical noise sources of the laser light is not present.  At this frequencies, the laser operates at or near the shot-noise limit.  In the 10\,Hz to
10\,kHz band where terrestrial GW detectors operate, 
technical noise masked and overwhelmed the observation of squeezing.  For example, acoustic, laser relaxation oscillatiion thermal and mechanical fluctuations can be many orders of magnitude larger than shot noise.  Until recently, it was not certain that a laser field could even be squeezed and matched to the
slow oscillation period of GWs.
Second, it was previously not known whether squeezed light was fully
compatible with other extremely sophisticated technologies employed
in GW detectors, such as signal-recycling.
Third, the technology to reliably produce stable and strong
squeezing with large squeeze factors was lacking.  Long term observation of strong squeezing was a technical challenge until recently.

These challenges have all been overcome in the past decade.  All the open questions have now been satisfactorily addressed.  This development is very timely since many known advanced classical interferometric techniques have almost been exhausted.  Many remaining classical improvements are becoming increasingly difficult and expensive) to implement.

\begin{figure}[ht]
  \vspace{-1mm}
  \includegraphics[width=8.6cm]{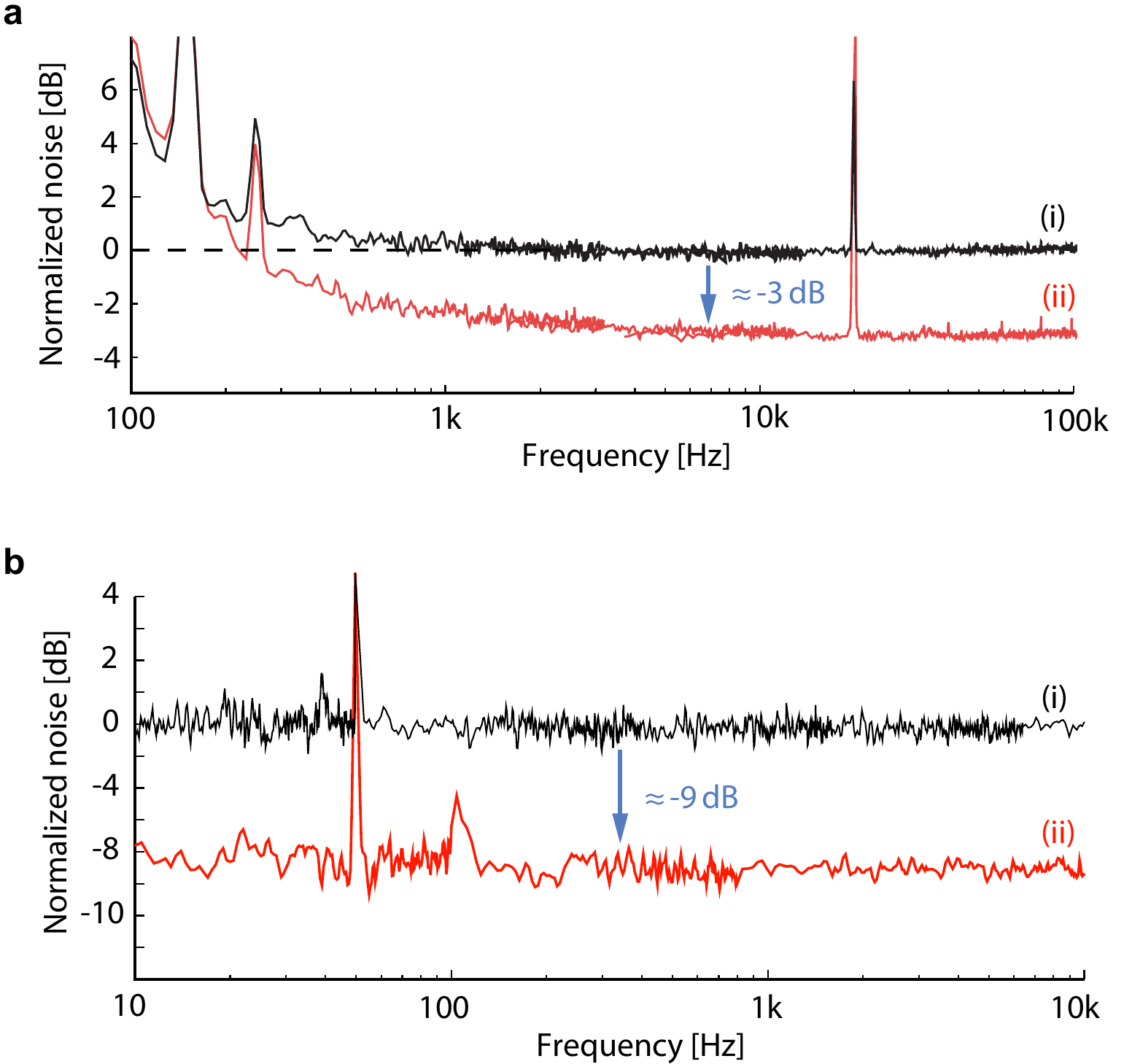}
  \vspace{-1mm}
\caption{{\textbf{Quantum noise squeezing}\; Both panels show the spectral analysis of measured noise powers without (i) and with `squeezing' (ii). The horizontal sections of traces (i) correspond to shot-noise, serving as reference levels (0\,dB), respectively. Top panel (a): The first audio-band squeezing down to about 200\,Hz was demonstrated by McKenzie \textit{et al.}\ in 2004 \cite{MGBWGML04}. Bottom (b): Current best performance of a squeezed light laser for GW detection shows an up to 9\,dB squeezed noise over the complete detection band of ground-based GW detectors \cite{Vahlbruch10AMALDI}.} 
}
\label{fig8}
\end{figure}

\textbf{Generation of squeezing in the audio-band}\; A major
breakthrough in achieving squeezing in the audio band was the
insight that the dominant noise at audio frequencies that
degrades squeezed light generation couples via the coherent laser
field that was used to control the length of the squeezed light
laser resonator, whereas noise coupling via the second harmonic
pump field is insignificant \cite{BSTBL02,Schnabel80kHz}. This led
to the first demonstration of audio-band squeezing at frequencies
down to 200\,Hz \cite{MGBWGML04}, see Fig.~\ref{fig8}\,a. There the length of the squeezing
resonator was stabilized without a bright control beam by using the phase sensitivity of the squeezing itself -- a technique known as quantum noise locking \cite{MMGLGGMM05}. Subsequently a coherent beam control scheme was invented \cite{Vahlb06} for simultaneous control of both the squeezing
resonator length and the squeezing angle \cite{GerryKnight04}. Shortly thereafter another noise source was
identified and mitigated, which allowed for squeezing of more than
6\,dB throughout the audio-band down to 1\,Hz \cite{Vahlb07}. This
noise source arose due to tiny numbers of photons that were
scattered from the main laser beam and rescattered into the audio
band squeezing mode after having experienced a frequency shift due
to vibrations and thermal expansions of potential scattering
surfaces, an effect known as \textit{parasitic interferences}. Since
bright laser beams cannot be completely avoided, the recipe for the
generation of audio-band squeezing turned out to be fourfold:
avoiding scattering by using ultra-clean super-polished optics,
avoiding rescattering by carefully blocking all residual faint beams
caused by imperfect anti-reflecting surfaces, reduce the
vibrationally and thermally excited motion of all mechanical parts
that could potentially act as a re-scattering surface and avoid pointing fluctuations \cite{MGLM07}.

\textbf{Compatibility of squeezing with other interferometer techniques}\; Current detectors achieve their exquisite
sensitivity to GWs due their kilometre-scale arm lengths, the
enormous light powers circulating in the enhancement resonators
(arm, power- and signal-recycling cavities), 
and sophisticated pendulum suspensions that isolate the test mass
mirrors from the environment (Fig.~\ref{fig3}). When these
techniques were developed, squeezing was not envisioned to become an
integrated part of such a system. Building on existing theoretical
work \cite{GLe87,HCCFVDS03}, a series of experimental demonstrations
of squeezed state injection into GW detectors were carried out.
These included compatibility with power recycling, with signal
recycling \cite{MSMBL02,Vahlb05}, and with the dynamical system of
suspended, quasi-free mirrors \cite{Goda08NatPhys,S-NaV08}.

\textbf{Generation of strong squeezing}\; Squeezing has significant
impact in quantum metrology if large squeezing factors can be
produced. Squeezing of 3\,dB improves the signal-to-noise ratio by a
factor of $\sqrt{2}$, equivalent to doubling the power of the
coherent laser input. Squeezing of 10\,dB corresponds to a ten-fold
power increase. Remarkably, the experimentally demonstrated
squeezing factors have virtually exploded in recent years
\cite{TYYF07,Vahlb08,CommentPolzik}, culminating in values
as large as 12.7\,dB \cite{Eberle10}. 
All the squeezing factors above
10\,dB were observed with monolithic resonators and at MHz
frequencies. However, reduced optical loss in non-monolithic resonators and a careful elimination of
parasitic interferences should in principle enable such factors
also in the GW band. 
An 8 to 10\,dB improvement based on strong squeezing seems
realistic for future GW detectors in their shot-noise limited band \cite{Eberle10}.

\textbf{The first squeezed light laser for GW detection}\; Based on
the previous achievements reviewed here, very recently, the first
squeezed light laser for the continuous operation in GW detectors was
designed and completed \cite{VahlbDiss,Vahlbruch10AMALDI}. Up to
9\,dB of squeezing over the entire GW detection band has been
demonstrated (Fig.~\ref{fig8}b). This laser produces squeezed vacuum
states and is fully controlled via co-propagating frequency-shifted bright control
beams. This 9\,dB squeezing factor is limited by technical effects:
The squeezing resonator has to have an adjustable air gap to allow
for an easy way to apply length control. The anti-reflection coated surface
in the resonator introduces additional loss and reduces the escape
efficiency. Moreover, a Faraday isolator has to be used in the
squeezed beam path in order to eliminate parasitic interferences.
This rotator produces a single pass photon loss of about 2\%. This
squeezed light source is designated for continuous operation in the
GEO600 GW detector. A squeezed light source based on a design that should have less sensitivity to retro-scattered light \cite{MGGLM06} is being prepared for deployment on one of the most sensitive
detectors, the 4\,km LIGO detector in Hanford, Washington.

\section{Future Directions}  

The final test of the squeezed light technology for GW astronomy can be carried out only in a (large scale) GW detector. During operation such a detector takes data 24 hours a day, 7 days a week, and future experiments will test appropriate electro-optical auto-alignment systems that continuously provide a high interference contrast between the extremely dim squeezed laser mode and the high-power laser mode at the interferometer's central beam splitter.
We are convinced that these experiments will be successful thereby establishing quantum metrology as a key technology for all next generations of GW detectors.\\
Since squeezed light builds on quantum correlations between photons, loss of photons reduces the squeezing effect. Future research therefore has to deal with a reduction of photon loss in GW detectors down to a few percent in order to be able to make use of the full potential of squeezed laser light. State of the art optical technologies are already able to provide such low loss. 
With a sufficiently reduced optical loss also the enhancement of the nonclassical noise suppression of squeezed light lasers is expedient again thereby preparing the ground for an even higher level of quantum noise reduction.\\ 
When targeting signal frequencies at which quantum shot noise is dominating squeezing will certainly be combined with further increased light powers. When targeting frequencies at which thermal noise and technical noise sources dominate, such as photon scattering, the squeezed light technology will be imbedded in a comprehensive low noise concept providing a new and versatile starting point. It will enable the combination of low shot noise, quantum non-demolition techniques, and the cryogenic operation of mirror test masses thereby helping to make GW astronomy reality. \\

~\\
\textbf{Acknowledgements} This research was supported by the
Australian Research Council, the National Science Foundation and the
Deutsche Forschungsgemeinschaft through SBF\,407, SFB\,TR7, and the Centre for Quantum
Engineering and Space-Time Research, QUEST. We apologize to all those colleagues
whose work was not cited because of space restrictions.

\vspace*{2mm}

\textbf{Competing Interests} The authors declare that they have no competing financial interests.

\vspace*{2mm}

\textbf{Correspondence} Correspondence and requests for materials should be addressed to
R. Schnabel (email: roman.schnabel@uni-hamburg.de).

\end{document}